# ARTICLE

# High magnetoresistance of hexagonal boron nitride- graphene heterostructure-based MTJ through excited-electron transmission

Halimah Harfah,*[a‡] Yusuf Wicaksono,[a‡] Gagus Ketut Sunnardianto,[c] Muhammad Aziz Majidi,[b] and Koichi Kusakabe[a]

This work presents an ab-initio study of a few-layers hexagonal boron nitride (hBN) and hBN-graphene heterostructure sandwiched between Ni(111) layers. The aim of this study is to understand the electron transmission process through the interface. Spin-polarized density functional theory calculations and transmission probability calculations were conducted on Ni(111)/$n$hBN/Ni(111) with $n$ = 2, 3, 4, and 5 as well as on Ni(111)/hBN-Gr-hBN/Ni(111). Slabs with magnetic alignment in an anti-parallel configuration (APC) and parallel configuration (PC) were considered. The $pd$-hybridizations at both the upper and lower interfaces between the Ni slabs and hBN were found to stabilize the system. The Ni/$n$hBN/Ni magnetic tunnel junction (MTJ) was found to exhibit a high tunneling magnetoresistance (TMR) ratio at $\sim 0.28$ eV for $n$ = 2 and $0.34$ eV for $n$ > 2, which are slightly higher than the Fermi energy. The observed shifting of this high TMR ratio originates from the transmission of electrons through the surface states of the $d_{z^2}$-orbital of Ni atoms at interfaces which are hybridized with the $p_z$-orbital of N atoms. In the case of $n$ > 2, the proximity effect causes an evanescent wave, contributing to decreasing transmission probability but increasing the TMR ratio. However, TMR ratio, as well as transmission probability, was found to be increased, by replacing the unhybridized hBN layer of the Ni/3hBN/Ni MTJ with graphene, thus becoming Ni/hBN-Gr-hBN/Ni. A TMR ratio as high as ~1200% was observed at an energy of 0.34 eV, which is higher than the Fermi energy. Furthermore, a design is proposed for a device based on a new reading mechanism using the high TMR observed just above the Fermi energy level.

## 1 Introduction

Magnetic tunnel junctions (MTJs) are one of the most important devices for spintronic applications. Many studies of MTJs have been undertaken[1–13], indicating the existence of a potential of the devices' application to logic devices, hard-drive magnetic read heads, and magnetic sensors[14–19]. The primary competition for improvement of MTJ is to attain a higher percentage of the tunneling magnetoresistance ratio (TMR). One of the vital points is its tunnel barrier. The common tunnel barrier used is MgO. Among various MTJs, CoFeB/MgO/CoFeB showed the highest TMR of $1100\%$ at 4.2K[20]. In order to downscale the device, retaining the transmission high and reducing the barrier thickness are essential. However, when this is realized, TMR can be reduced to $55\%$ due to the presence of uncontrollable defects in the MgO tunnel barrier[21].

Recently, 2D materials have been examined for the non-magnetic spacers in current-perpendicular to-plane (CPP) MTJs. Currently, the performance on the TMR ratio is still relatively low[22]. However, one atomic thickness material, such as graphene and hexagonal Boron Nitride (hBN), owing to their extraordinary properties, is expected to arise a unique characteristic of MTJ and a new device mechanism that could not be found in the conventional one. The chemical and physical interaction at the interfaces between graphene or hBN with the ferromagnetic layer are interesting and can be further developed. For instance, when graphene is realized as a tunnel barrier of MTJ, small magnetoresistance (MR) was found when the CPP is considered[2,10,11,22–29]. However, a recent theoretical study suggested that when the Ni/graphene/Ni system is considered, a controllable gapped-Dirac cone can be realized, leading to a high current-in-plane (CIP) magnetoresistance ratio[30]. A similar case is also found in our previous study on monolayer hBN. A novel functionality of spin-current control through cross-correlation properties of the bi-stable state at Ni/hBN/Ni junction interfaces was recently discovered, which could not be obtained in conventional MTJs[31].

A van der Waals heterostructure of graphene-hBN has recently become huge interest due to its unique properties. A fabrication of graphene-hBN van der Waals heterostructure has been further developing, and the creation of the graphene-hBN heterostructure has been possible[32]. Previous experimental study shows the use of graphene-hBN heterostructure as a tunnel barrier of 2D materials based MTJ and found an

[a.] *Graduate School of Engineering Science, Osaka University, 1-3 Machikaneyama-cho, Toyonaka, Osaka 560-8531, Japan.*
[b.] *Department of Physics, Faculty of Mathematics and Natural Science, Universitas Indonesia, Kampus UI Depok, Depok, Jawa Barat 16424, Indonesia.*
[c.] *Research Center for Physics, Indonesian Institute of Sciences, Kawasan Puspiptek Serpong, Tangerang Selatan, 15314, Banten, Indonesia.*
\* Corresponding author; E-mail: u268634c@ecs.osaka-u.ac.jp
‡ These authors contributed equally to this work

increasing performance compared with that of graphene or hBN based MTJs, as well as a unique inverted signal of the spin valve, have been found[33]. The unique chemical and physical properties are expected between the interface and insulator barrier. Therefore, a further theoretical investigation is required to attain a full understanding of the MTJ properties.

In this paper, a comprehensive theoretical study on a Ni/hBN- Gr-hBN/Ni MTJ is presented. It was found that this MTJ exhibits a TMR ratio as high as ~1200% at energy slightly higher than the Fermi Energy, and it is envisaged that this material system will have a new reading process mechanism. To understand the unique properties at the interfaces of the Ni/hBN-Gr-hBN/Ni MTJ, an investigation was also conducted on Ni/$n$hBN/Ni MTJ with $n = 2, 3, 4,$ and 5. Also, in this case, it was found that the high TMR ratio does not occur at the Fermi energy level but rather at slightly higher energy, namely 0.28 eV for $n = 2$ and 0.34 eV for $n > 2$. The shifting of the high TMR ratio originates from the transmission of electrons through the surface state of the $d_{z^2}$-orbital of the Ni atoms at the interfaces, which hybridize with the $p_z$-orbital of the N atoms. When $n > 2$ is considered, a proximity effect takes place, which results in an evanescent wave, and contributes to the change of transmission probability, hence leading to a change in the TMR ratio as well. The evanescent wave still exists at $E - E_F = 0.34$ eV but is rather weak due to the proximity effect of $d_{z^2}$-orbital of Ni works on $p_z$-orbital of Boron, which is unoccupied. The role of the graphene layer in Ni/hBN-Gr-hBN/Ni MTJ becomes a key point to strengthen the proximity effect of the $d_{z^2}$-orbital inside the insulator barrier, resulting high TMR.

## 2 Computational Method

Spin-polarized plane wave-based density functional theory (DFT) calculations were performed using the Quantum ESPRESSO package[34,35]. A revised Perdew-Burke-Ernzerhof (PBE) functional for a densely packed solid surface, called the PBESol functional[36], as well as ultrasoft pseudopotentials[37], within the generalized gradient approximation (GGA) were used to describe the electron-ion interaction. A kinetic energy cut-off of 50 Ry was used for the wavefunctions to reach a good convergence calculation. Since adopting an appropriate $k$-point grid can result in the convergence of the total energy calculation in this system, a $k$-point grid of 24 × 24 × 1 was chosen for all calculations. Furthermore, the effect of the van der Waals interactions on the electronic structures was included by applying the most recent and accurate DFT-D3 corrections[38]. Taking into account the van der Waals interactions in the DFT calculations, the results are in good agreement with experimental values, especially regarding structures.

The ballistic transmission probability calculations were performed using the Landauer-Buttiker formalism[39,40]. The left lead, right lead, and scatterer regions were considered in the model calculation, as shown in Fig. 1. Ni(111) was considered for the left and right leads to reduce the computational cost without neglecting any important physics. However, the use of Au(111) as a lead carrying normal input/output currents for the Ni/$n$hBN/Ni and Ni/hBN-Gr-hBN/Ni structure would be expected in an actual device. The current flow from the left to the right lead through the scatterer can be expressed as follows:

$$I = \frac{e}{h} \int T(E)[f_L(E) - f_R(E)]dE, \quad (1)$$

where $f_L(E)$ $(f_R(E))$ are the right-(left-)moving electrons injected from the left (right) lead, respectively. Additionally, the ballistic transmission $T$ as a function of the energy $E$ is described as

$$T(E) = \sum_{\mathbf{k}_\parallel} \sum_{i,j} T_{i,j}(\mathbf{k}_\parallel, E), \quad (2)$$

where $T_{i,j}(\mathbf{k}_\parallel, E)$ is the probability of electrons with energy $E$ and state. $T(E)$ is then obtained by summing over the 2D Brillouin momentum $k$ to move from the $i$-th Bloch state to the $j$-th Bloch zone and over all incoming-outcoming states. In the present study, the transmission probability calculations were performed using the PWCOND[41] module of the Quantum ESPRESSO software. A perpendicular $k$-point of 50 × 50 respect to the transmission direction is considered to get a good accuracy of transmission probability.

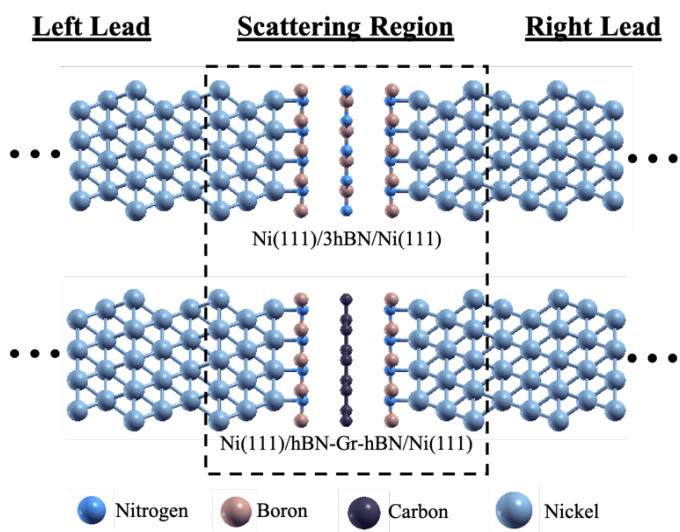

Fig. 1 Side view of the supercell of Ni(111)/3hBN/Ni(111) (as representative of few-layer hBN MTJ) and Ni(111)/hBN-Gr-hBN/Ni(111) used to represent the scattering region and lead corresponding to model calculation for transmission probability calculation.

The PWCOND module calculates the transmission probability at zero temperature and for an infinitely small voltage, in such a way that the conductance of the system can be calculated as follows:

$$G = \frac{1}{\delta V} = \frac{e^2}{h}T(E_F). \quad (3)$$

Finally, a magnetoresistance ratio calculation was conducted based on the Julliere model[42] by including the difference between the conductance in the APC and PC states, and then dividing it by the conductance in the APC state, according to:

$$TMR = \frac{G_P - G_{AP}}{G_{AP}}. \quad (4)$$

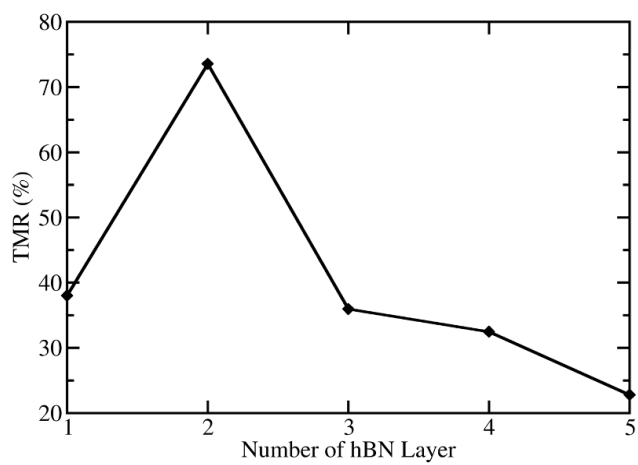

Fig. 2 The TMR ratio vs the number of hBN layers as tunnel barrier in Ni/nhBN/Ni with n = 1, 2, 3, 4, and 5. The values are taken at the zero-bias limit.

## 3 Results and Discussions

### 3.1 Transmission Mechanism of Ni/hBN/Ni MTJs with Different Numbers of hBN Layers

The pd-hybridizations at both the upper and lower interface between the Ni slabs and hBN stabilize the system. This relevant chemical property between Ni and hBN also is at the basis of the unique electron transmission phenomenon appearing in the hBN-based junctions. Large amount of evidence supporting this idea is presented in the following.

Fig. 2 shows the TMR ratio of hBN-based MTJs as a function of the number of hBN layers. The TMR value was taken at the Fermi energy, namely, the zero-bias limit. The TMR ratio profile is in good agreement with the previous theoretical study[43] and is consistent with several experimental studies[22].

When a double-hBN layer having two stacked hBN planes (2hBN) is considered as a tunnel barrier, the TMR ratio increases compared with a monolayer of hBN. Such increase is due to a difference in the transmission process of the electrons. When a monolayer of hBN as is considered as a tunnel barrier, due to $pd$-hybridizations from both upper and lower Ni slabs coupled with N atoms, charge transfer occurs leading to the hBN layer becoming metallic[31]. Thus, the propagating-wave electrons become the dominant contribution for the transmission between the two Ni(111) electrodes.

In addition 2hBN is used as a tunnel barrier, due to the weak van der Waals interaction between the two hBN layers, electrons are transmitted through the tunnel barrier via the Ni(111) surface state at the interface. This transmission process can be observed from the transmission probability profile of the Ni/2hBN/Ni system in Fig. 3(a). It is shown that the high transmission peak in the minority (majority) spin channel appears when the system in the PC state is shifted towards an energy higher (lower) than Fermi energy by 0.3 eV (−0.5 eV).

On the other hand, the local density of states (LDOS) of hBN at the interfaces in Fig. 3(b) shows newly created states at the insulator gap of hBN. Furthermore, the LDOS of those states shows a correlation with the profile of the transmission

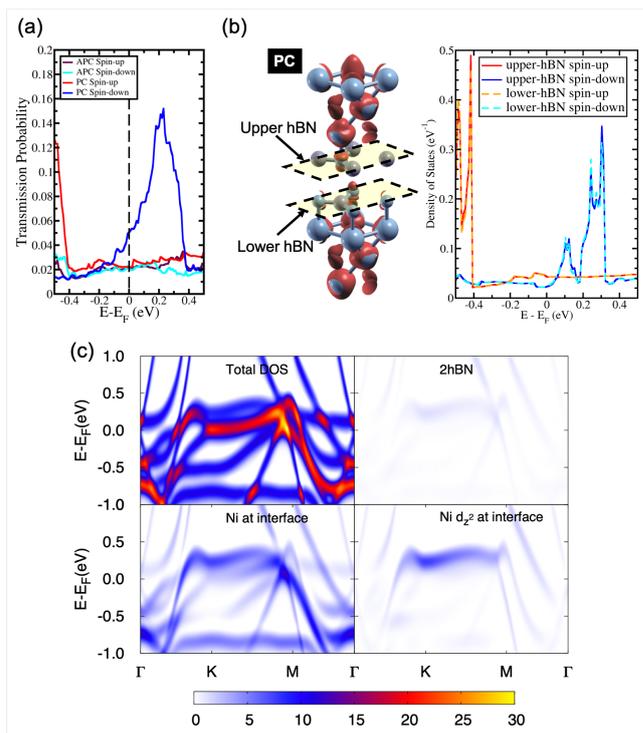

Fig. 3 (a) The transmission probability of Ni/2hBN/Ni , (b) The spin- charge density mapping (red color represent spin-up charge density mapping) and LDOS of 2hBN insulator barrier in Ni/2hBN/Ni system in PC state, (c) The projected bandstructure of Ni/2hBN/Ni probability of the system, whereby an increase in the density of states at an energy higher (lower) than the Fermi energy in the spin minority (majority) channel corresponds to a high transmission peak in the spin minority (majority) channel. Therefore, these states represent the dominant contribution to transmission of electrons through the insulator barrier.

The projected band structure in Fig. 3(c) shows that these states are originated from the Ni(111) surface state with major $d$-components. The high transmission peak observed at an energy higher (lower) than the Fermi energy in the spin minority (majority) channel corresponds to the $d_{z^2}$-orbital of Ni(111) at the interfaces which hybridize with the $p_z$-orbital of the N atoms. Simultaneously, the flat and broad states in Fig. 3(b) correspond to the s-orbital of Ni(111) at the interface. The relatively larger components in the $d_{z^2}$-orbitals compared with the $p_z$-orbitals suggest that the wavefunction is indeed in a tunnel regime, where 2hBN behaves as a potential barrier for the Ni d-electrons.

When the number of hBN layers is increased further, e.g., 3hBN, the contribution from these surface states at the Fermi energy is quenched. The surface states of Ni(111), which are derived from the $s$-orbital of the system with a number of hBN layers higher than-two, becomes weaker around the Fermi energy, as shown in Fig. 4(b). This quenching leads to a drop in the transmission probability of electrons in the spin minority channel of the PC state, which becomes approximately equivalent to the transmission probability of electrons in the APC state as shown in Fig. 4(a). This result yields the TMR ratio of the system, which becomes lower than that observed for Ni/2hBN/Ni. Furthermore, when the number of hBN layers is

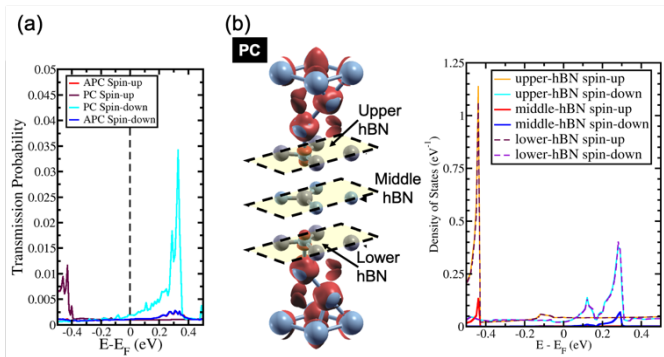

Fig. 4 (a) The transmission probability of Ni/3hBN/Ni and (b) The spin charge density mapping of Ni/3hBN/Ni in PC state (red color represent spin-up charge density mapping) and LDOS of 3hBN insulator barrier in Ni/3hBN/Ni system in PC state.

further increased to 4hBN and 5hBN, a monotonic decrease in the TMR ratio with respect to the increasing number of hBN layers is expected just around the Fermi energy.

### 3.2 High Transmission Magnetoresistance on the Excited State and Influence of the Proximity Effect

In the last section, the presence of a peak in the transmission probability was discussed, as shown in Fig. 3 and Fig. 4. The high transmission peak of electrons observed at an energy slightly higher than Fermi energy is still present in the spin minority channel upon increasing further the number of hBN layers.

From the transmission probability of electrons for Ni/$n$hBN/Ni with $n$ = 4 and 5, a high and increasing TMR ratio is observed at an energy higher than the Fermi energy. The peak of the transmission probability is located at energy by 0.3 eV higher than the Fermi energy, as shown in Fig. 5(a) and (c).

The LDOS of the unhybridized hBN layer for the Ni/$n$hBN/Ni system with $n$ = 4 and 5 are shown in Fig. 5(b) and (d), respectively. The typical behavior corresponding to the presence of the proximity effect, which causes an evanescent wave, or a dumping mode inside hBN, is found as a lower LDOS in the center hBN layer than other hBN layers closer to Ni slabs. At an energy close to the Fermi energy, since the energy gap of hBN does exist, no real propagating modes in the hBN slab is expected. However, the effect is rather weak since the proximity effect of $d_{z^2}$ acts on the $p_z$ orbital of B, which is unoccupied. Thus, for the case of Ni/$n$hBN/Ni with $n>2$, the proximity effect becomes the main contribution for the transmission through the hBN tunnel barrier

This tunneling behavior causes an interesting filtering effect. When energy range is 0.3 eV higher than the Fermi energy, the minor-spin channel exhibits a high transmission probability. By contrast, for the major-spin channel, the transmission is largely reduced. The high transmission for the minor-spin channel originates from the $d$-orbital nature, since a large $pd$-hybridization causes a spin-split LDOS with the large peak at the corresponding energy. Thus, except for the $d$ channel, a lowered transmission can only be expected. Therefore, the major-spin component shows a reduced transmission. As a result, the ratio of the minority-spin transmission and majority-spin transmission should become more prominent as $n$ is increased.

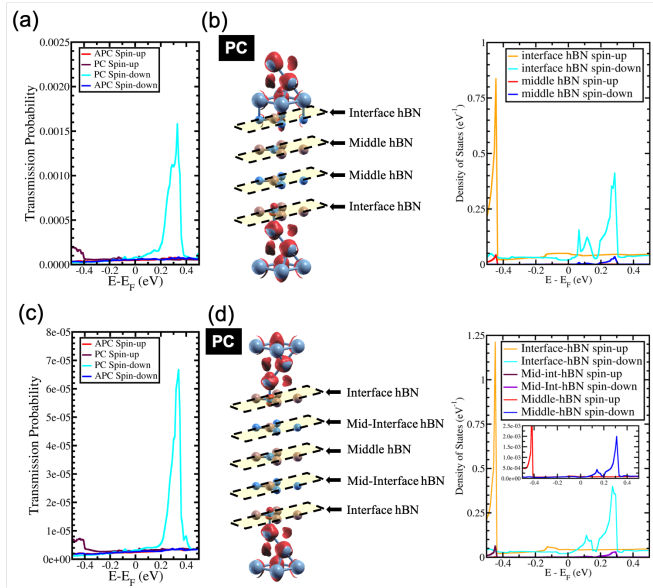

Fig. 5 (a) The transmission probability of Ni/4hBN/Ni, (b) The charge density mapping of Ni/4hBN/Ni (red color represent spin-up charge den- sity mapping) and the LDOS of 4hBN insulator barrier of Ni/4hBN/Ni sys- tem in PC state, (c) The transmission probability of Ni/5hBN/Ni, and (d) The charge density mapping of Ni/5hBN/Ni (red color represent spin- up charge density mapping) and the LDOS of 5hBN insulator barrier of Ni/5hBN/Ni system in PC state (insert: magnified LDOS of middle hBN of Ni/5hBN/Ni system).

This large transmission probability is found only when a parallel-spin configuration is selected with the pd-hybridization occurring on both sides of hBN. In the case of the APC, a blocking behavior can be observed. As a result, the TMR ratio also becomes significant at a sharp LDOS position of 0.3 eV above the Fermi energy.

Interestingly, the effect of the proximity can be magnified when the hBN layer is replaced with the graphene layer. For example, by replacing the unhybridized hBN layer of the Ni/3hBN/Ni MTJ with graphene, thus becoming Ni/hBN-Gr-hBN/Ni. Fig. 6(b) shows the LDOS of the hBN-Gr-hBN insulator barrier, which shows a higher electronic density of states on graphene at 0.34 eV. Thus, these states lead to a higher transmission probability at the corresponding energy. As shown in Fig. 6(a) high magnetoresistance ratio of ~1200% can be observed for the Ni/hBN-Gr-hBN/Ni MTJ at energy 0.34 eV. This high performance and unique characteristics of the Ni/hBN-Gr-hBN/Ni MTJ could provide novel functionalities, such as optically induced MTJs, which are introduced in the following section.

### 3.3 Proposed Design and Mechanism of Optically Induced MTJs

The proposed idea of using a Ni/hBN-Gr-hBN/Ni MTJ as an optically induced MTJ is shown in Fig. 7. The process of reading and writing data in a memory using light irradiation is here considered. For the writing process, as discussed in previous study, an optical demagnetization using a circularly polarized femtosecond or sub-picosecond laser and field-assisted magnetic switching is capable of changing the magnetic orientation of the free Ni(111) ferromagnetic layer[44]. The high performance and unique characteristics of Ni/hBN-Gr-hBN/Ni MTJ are primarily used for the reading process. Two Au

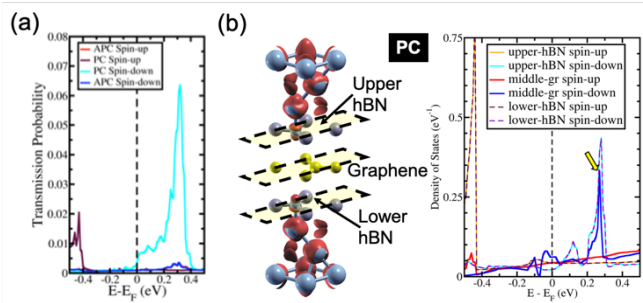

Fig. 6 (a) The transmission probability of Ni/hBN-Gr-hBN/Ni and (b) The spin charge density mapping of Ni/hBN-Gr-hBN/Ni (red color represent spin-up charge density mapping) and LDOS of hBN-Gr-hBN insulator barrier of Ni/hBN-Gr-hBN/Ni system in PC state.

electrodes were used to flow the current into the proposed MTJ device. A small bias voltage was applied. At first, the current flows from the Au electrode to the transparent electrode on the top of the Ni/hBN-Gr-hBN/Ni MTJ. Without any light irradiation onto the system, the reading process is not able to be optimally carried out, since whether the current passing- through the MTJ is relatively small in both APC and PC states. This is due to the fact that, at Fermi energy, the electron transmission for the APC and PC state is low as explained in the previous section. The reading conditions 1 and 2 are illustrated in Fig. 7(b). When linearly polarized light, e.g., infrared or visible light, is used to excite the electrons of the upper Ni slabs to the energy 0.34 eV higher than the Fermi energy, the optimal reading process is observed. This condition (#3) is illustrated in Fig. 7(b), where it is shown that, a high transmission occurs in the PC state resulting in the flow of current through the MTJ towards the lower Au electrode. In addition, condition 4 shows that the MTJ is in the APC state, thus explaining the observed low transmission.

## 4  Conclusions

Investigation into Ni/$n$hBN/Ni MTJs were conducted by increasing the number of hBN layers in the tunnel barrier. Owing to the electrons transmission through the surface states, an increasing TMR ratio was observed when considering 2hBN as the tunnel barrier compared with the case of a monolayer hBN. However, a monotonic decrease in TMR was found when more than two hBN layers were considered. The reason for this behavior is due to the quenched surface states of Ni(111) at Fermi energy, which occur at the insulator gap of hBN.

On the other hand, it was found that the Ni/$n$hBN/Ni MTJ exhibits a slight shift of the highest transmission peak towards an energy higher than the Fermi energy. This result comes from the electrons transmission through Ni $d_{z^2}$ which hybridized with N $p_z$. Interestingly, A high and increasing TMR ratio was observed at the energy where the high transmission peak is located. These results are due to the proximity effect of the unhybridized hBN layer. The effect of the proximity can be magnified when the hBN layer is replaced with the graphene layer. When Ni/3hBN/Ni is considered and the unhybridized hBN layer in Ni/3hBN/Ni is

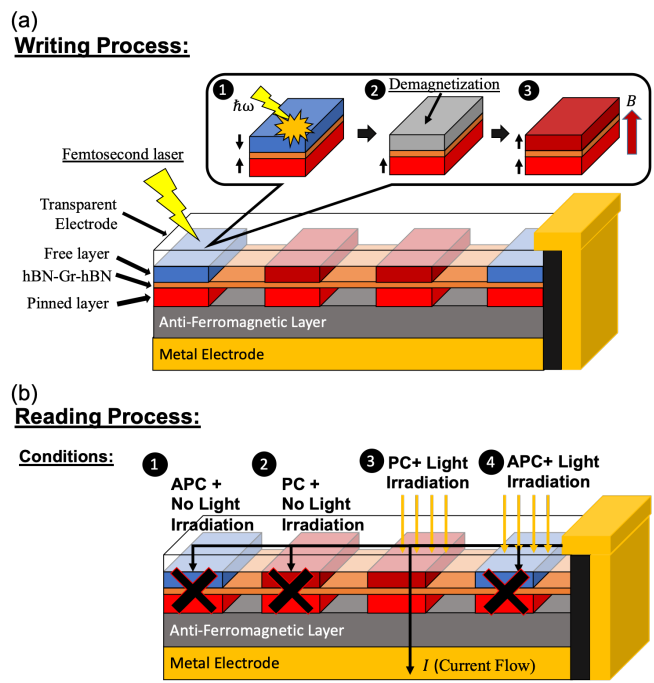

Fig. 7 (a) The writing and (b) the reading process of Ni/hBN-Gr-hBN/Ni proposed MTJ device with light irradiation.

replaced with a graphene layer, the stronger proximity effect becomes evident through a higher electronic density of states at the corresponding energy for the LDOS of graphene. Thus, these states lead to a higher transmission probability at the corresponding energy. A high magnetoresistance ratio of $\sim 1200\%$ was observed for the Ni/hBN-Gr-hBN/Ni MTJ at an energy of $0.34$ eV.

The high performance and unique characteristics of this Ni/hBN-Gr-hBN/Ni MTJ provide the possibility to exploit a novel device functionality, namely, that of an optically induced MTJ. The process of reading and writing in this proposed MTJ is expected to be conducted by light irradiation. Optical demagnetization and field-assisted magnetic switching is expected to be able to change the magnetic orientation of the free Ni(111) ferromagnetic layer. This process would represent the writing process in the proposed device. On the other hand, the unique characteristics of the Ni/hBN-Gr-hBN/Ni MTJ would mainly contribute to the reading process. The Linear polarized of light was applied to induce the transmission to occur at energy higher than Fermi energy by $0.34$ eV. This process was exploited to read the magnetic alignment of the Ni/hBN-Gr-hBN/Ni MTJ.

## Author Contributions

Conceptualization was done by H.H. and Y.W. H.H. performed the transmission probability and found high TMR at energy higher than Fermi energy. The charge density mapping was performed by H.H. Y.W. performed LDOS and projected bandstructure to understand the transmission mechanism. The new mechanism of writing and reading process was proposed by Y.W. K. K. performed a supervision on the research and proposed a detail explanation on proximity effect. Y.W. and H.H. wrote the original draft of manuscript. K. K., G. K. S., and M. A. M. wrote and edited the manuscript. All the authors reviewed the manuscript.

## Conflicts of interest

"There are no conflicts to declare".

## Acknowledgements

The calculations were performed at the computer centers of Kyushu University. This work was partly supported by JSPS KAKENHI Grant No.JP26400357, JP16H00914 in Science of Atomic Layers, JP18K03456, and 20J22909 in Grant-in-aid for Young Scientist. Y. W. gratefully acknowledge fellowship support from Japan Society for the Promotion of Science (JSPS). H. H. grate- fully acknowledge scholarship support from the Japan International Cooperation Agency (JICA) within the "Innovative Asia" Program, ID Number D1805252.

## Notes and references